\documentclass[12pt]{iopart}
\usepackage{graphicx}

\usepackage{csquotes}
\usepackage{lipsum}
\usepackage{float}
\usepackage{xcolor}
\usepackage{subcaption}
\usepackage{gensymb}
\usepackage{multirow}

\usepackage[backend=biber,style=authortitle,bibstyle=numeric, citestyle=phys,doi=false,url=false,isbn=false, maxnames = 20]{biblatex}

\begin{document}


\title[Deformable Scintillation Dosimeter II : Dose and DVF measurements]{Deformable Scintillation Dosimeter II: Real-Time Simultaneous Measurements of Dose and Tracking of Deformation Vector Fields}

\author{E Cloutier$^{1,2}$, L Beaulieu$^{1,2}$ ,L Archambault$^{1,2}$
\footnote{Present address:
Department of Physics, Université Laval, Quebec,
QC BS8 1TS, Canada.}
\address{$^1$ Physics, physical engineering and optics department and Cancer Research Center, Universite Laval, Quebec, Canada.}
\address{$^2$ CHU de Quebec - Université Laval, CHU de Quebec, Quebec, Canada}}

\begin{abstract}
Anatomical motion and deformation pose challenges to the understanding of the delivered dose distribution during radiotherapy treatments. Hence, deformable image registration (DIR) algorithms are increasingly used to map contours and dose distributions from one image set to another. However, the lack of validation tools slows their clinical adoption, despite their commercial availability. This work presents a novel water-equivalent deformable dosimeter that simultaneously measures the dose distribution and tracks deformation vector fields (DVF). The dosimeter in made of an array of 19 scintillating fiber detectors embedded in a cylindrical elastomer matrix. It is imaged by two pairs of stereoscopic cameras tracking the position and angulation of the scintillators, while measuring the dose. The resulting system provides a precision of 0.3 mm on DVF measurements. The dosimeter was irradiated with 5$\times$3, 4$\times$3 and 3$\times$3 cm$^2$ 6 MV photon beams in both fixed and deformed conditions. The measured DVF was compared to the one computed with a DIR algorithm (Plastimatch). The deviations between the computed and measured DVFs was below 1.5 mm. As for dose measurements, the dosimeter acquired the dose distribution in fixed and deformed conditions within 1\% of the treatment planning system calculation and complementary dose validation using the Hyperscint dosimetry system. Using the demonstrated qualities of scintillating detectors, we developed a real-time, water-equivalent deformable dosimeter. Given it's sensor tracking position precision and dose measurements accuracy, the developed detector is a promising tools for the validation of DIR algorithms as well as dose distribution measurements under fixed and deformed conditions.

\end{abstract}

\maketitle


\section{Introduction}
Advances in modern radiotherapy treatment techniques have led to the advent of complex personalized treatment plans aimed at maximizing the dose delivered to the tumor while minimizing the dose delivered to surrounding tissues. Treatments plans are personalized to the patient's anatomy, resulting in dose gradients close to the target. However, over the course of treatments, the patient's anatomy may be deformed and/or change in volume. These anatomical variations challenge the understanding of the cumulative dose delivered throughout the course of radiotherapy treatments \cite{yeo_is_2012}. Hence, deformable image registration (DIR) algorithms are increasingly used in the clinics to either map organ contours or dose distribution from one image set to another\cite{yuen_international_2020}. However, in low contrast tissues, the high number of degrees of freedom of these algorithms can lead to inaccuracies in the computed deformation vector field (DVF) \cite{taylor_comment_2013,schultheiss_it_2012,kirby_need_2013}. Using those DVFs would result in incorrect voxel pairing, leading to errors in dose accumulation. Thus, the American Association of Physicist in Medicine Task Group 132 on the use of image registration algorithm in radiotherapy (TG-132) recommends that end-to-end tests should be performed using quality assurance (QA) phantoms prior to the implementation of these systems in the clinics \cite{brock_use_2017}. In spite of the these recommendations, the definition of a patient-specific gold standard DIR validation tool remains an open issue \cite{paganelli_patient-specific_2018}. Amongst the proposed validation tools, physical phantoms benefit from their ability to test the entire registration process, from the image acquisition to the registration itself. 

Deformable dosimetric gels have shown potential in measuring three-dimensional dose distributions delivered to deformable targets \cite{yeo_novel_2012, niu_novel_2012, deene_flexydos3d:_2015, matrosic_deformable_2019}. These water-equivalent gels demonstrated robust reproducibility and spatial resolution up to 1 mm \cite{juang_preliminary_2013}. However, they are integrating dosimeters and thus can only provide information on the cumulative dose deposited. Some anthropomorphic phantoms were also developed using landmarks to measure solely the deformation, not the dose \cite{kirby_two-dimensional_2011, serban_deformable_2008-1}. Some deformable phantoms were further developed with enclosures to insert  ion chambers, radiochromic films or MOSFETs, for dose measurements \cite{liao_anthropomorphic_2017, perrin_anthropomorphic_2017, gholampourkash-def}. However, the non-water equivalence of these dosimeters limits the practical number of simultaneous measurement points as some detectors can disturb the dose deposition pattern. Moreover, the contrast associated with these detectors may bias DIR validation in homogeneous mediums since it could be interpreted as fiducial markers in the images by the algorithms. 

On the other hand, work on volumetric scintillation detectors has shown the feasibility of real-time dose measurements over whole 2D and 3D volumes \cite{rilling_tomographicbased_2020, rilling_simulating_nodate, goulet_novel_2014,  kroll_preliminary_2013, guillot_toward_2010, kirov_three-dimensional_2005, kirov_new_2000}. Those systems provide millimeter resolution and water-equivalent measurements, but was limited to fixed measurements. As scintillators possess essential dosimetric qualities \cite{beaulieu_review_2016}, they may constitute an ideal candidate for the sensitive volume of a volumetric deformable dosimeter \cite{part1-submittedpaper}. Such a dosimeter could be suited for both the challenges of motion management and advanced radiotherapy modalities. 

This work presents the development of a novel scintillator-based deformable detector that simultaneously measures the dose distribution and tracks deformation vector fields at 19 positions.  

\section{Methods}
\begin{figure}
    \centering
    \includegraphics[width = 0.9\textwidth]{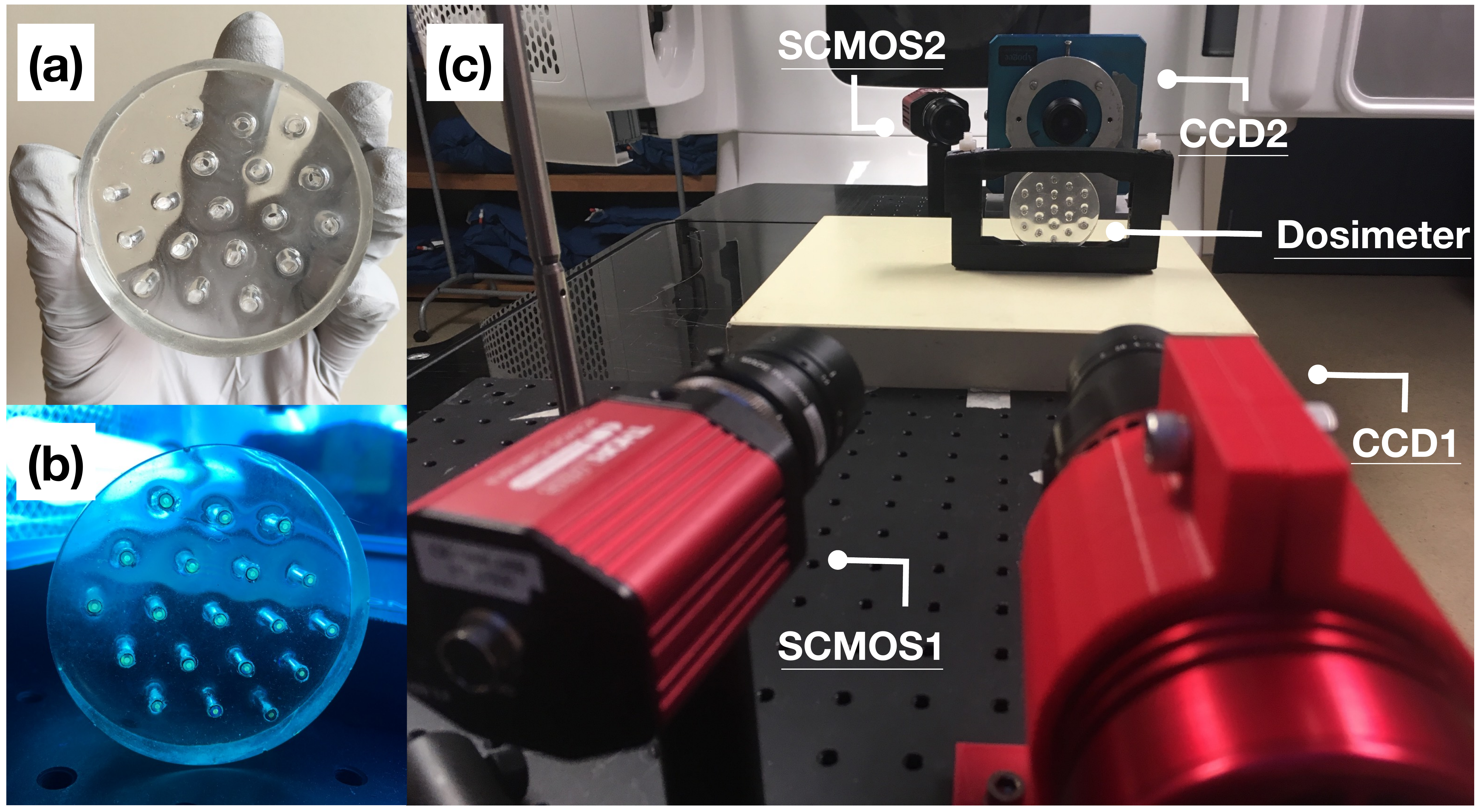}
    \caption{Representation of the developed dosimeter and its experimental set-up. (a) Clear deformable elastomer matrix. (b) Dosimeter composed of the elastomer matrix with 19 scintillating fibers embedded. (c) Experimental irradiation setup }
    \label{fig:methods}
\end{figure}{}

\subsection{Dosimeter description}

The dosimeter consists of 19 scintillators embedded in a clear, water-equivalent elastomer (figure \ref{fig:methods}b). The elastomer (Clearflex30: Smooth-On, Macongie, USA) was cast in a silicone cylindrical mold (diameter: 6 cm, thickness: 1.2 cm) and the compound was degassed to ensure an optimal transparency of the bulk.  Physical properties of the elastomer are listed in table \ref{tab:properties}.

\begin{table}[ht]
\caption{Physical properties of the clear plastic matrix provided by the manufacturer.}
\label{tab:properties}
\footnotesize\rm
\centering
\begin{tabular}{@{}lllll}
\br
Density & Refractive index & Tensile strength & Elongation at break & Shore hardness\\

[g/cm$^3$] & [-] & [psi] & [\%] & [A]\\
\mr
1.03 & 1.486 & 725 & 675 & 30 \\ 
\br
\end{tabular}
\end{table}

After pouring the gel, 19 polyethylene terephthalate (PET) tubes (Nordson medical, Salem, USA) were inserted in the elastomer guided by a 3D printed template. Once the elastomer set, the holder was removed, leaving an array of 19 hollow tubes in the cylindrical gel matrix, as can be seen on figure \ref{fig:methods}a. The hollow tubes have an internal diameter of 2.44 $\pm$ 0.03 mm to allow the insertion of the cylindrical scintillators assembly. 
Scheme and descriptions of the scintillators assembly can be found in table \ref{tab:scintillator_properties}. The scintillators consists of 1 mm diameter BCF-60 green scintillating fibers (Saint-Gobain Crystal, Hiram, OH, USA), inserted in a PET tubing (internal diameter of 1.1 $\pm$ 0.03 mm and external diameter of 2.16 $\pm$ 0.03 mm) covered with an opaque polyester heat-shrinking cladding (Nordson medical, Salem, USA). The scintillators were cut to a length of 1.2 cm to match the thickness of the elastomer matrix, and polished on both ends. 

\begin{table}[ht]
\caption{Scheme and description of the scintillators assembly. On the drawing, the dashed line delimits the tube attached to the elastomer and the scintillator assembly that can be inserted or removed. }
\label{tab:scintillator_properties}
\begin{tabular}{ll|lcc}
\br
 &   & Material                 & Internal diameter & Outer diameter \\
 & & &   [mm] &  [mm] \\
 \mr
\multirow{4}{*}{\includegraphics[width = 0.23\textwidth]{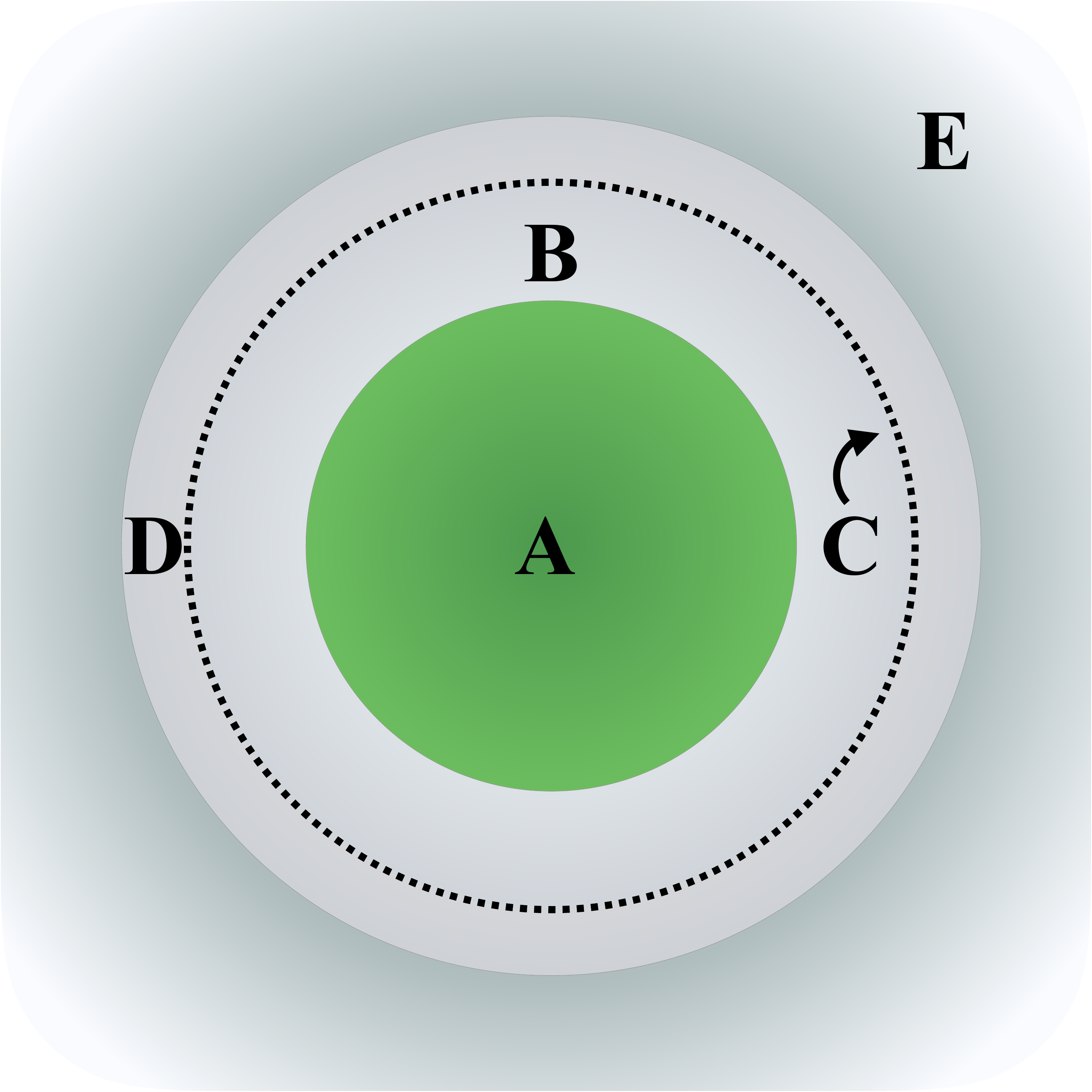} } &  &  A: BCF-60 scintillator  & -  & 1.00 \\
&  & B: PET tubing & 1.1   & 2.16     \\  
&  & C: Polyester tubing & 2.16     & 2.25 \\ & & D: PET tubing & 2.44  & 2.69 \\
& & E : Elastomer & -  & 60 \\
\\
\\
\br
\end{tabular}
\end{table}

A 1 cm vertical compression was applied to the dosimeter in the antero-posterior direction. The dosimeter was inserted between two plastic plates distant by 6 cm (fixed) and 5 cm (deformed state). The plates were brought closer with two tighten nylon screws (figure \ref{fig:methods}c).

\subsection{Detector assembly}

\begin{figure}
    \centering
    \includegraphics[width = 0.7\textwidth]{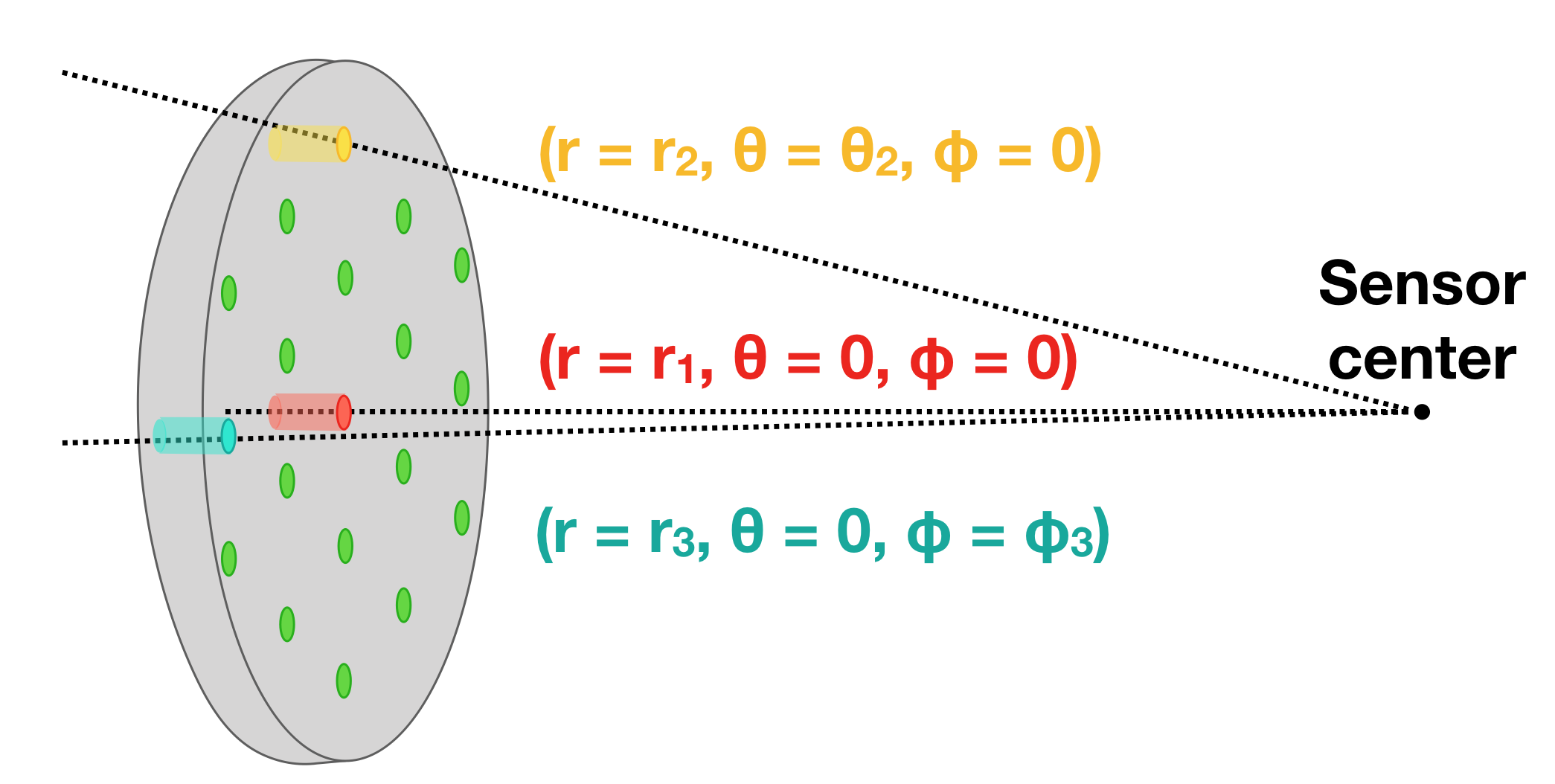}
    \caption{Schematic representation of angular and distal corrections. Angles and distances are measured using the vector connecting the tips of the scintillators to the CCD's sensor center.}
    \label{fig:angle-scheme}
\end{figure}

 The dosimeter was simultaneously imaged by 4 cameras as depicted on figure \ref{fig:methods}c. As scintillating fibers emit light in proportion to the dose deposited in their volume, collecting this signal provides information on the dose delivered as well as the scintillators location in the phantom. The cameras were arranged to form two facing stereoscopic pairs. Therefore, the setup enables the 3D position tracking of both ends of each scintillator. All cameras were coupled to 12 mm focal length lenses (F/\# = 16). The frame rate was set to 1s. Each pair consists of one CCD camera and one sCMOS. First, a cooled CCD camera (Atik 414EX; Atik Cameras, Norwich, United Kingdom) imaged the dosimeter and carried the radiometry analysis for dose measurements. The CCD1 was positioned 35 cm from the dosimeter. Another CCD (Alta U2000, Andor Technology, Belfast, United Kingdom) was placed on the other side of the dosimeter, facing the CCD1. Then, two sCMOS cameras (Quantalux, Thorlabs, Newton, USA) were paired to the CCDs to provide additional spatial information on the set-up. Since the deformation of the dosimeter leads to displacement and angle change of the scintillators, those movements result in signal variations, not related to the dose deposited. Those need to be corrected for \cite{part1-submittedpaper}. The stereoscopic pairs provide a complete 3D position tracking that makes possible angular and distal corrections. Vignetting corrections are also applied to each pixel, using a $\cos^4 (\theta_{(i,j)} )$ fit \cite{robertson_optical_2014}. The stereoscopic pair was calibrated using a (15$\times$10) grid chessboard pattern and a calibration algorithm inspired by Zhang from the OpenCV python library version 3.4.2 \cite{zhang_flexible_2000, opencv_library}. The scintillation signal was corrected according to their angle and distance from the CCD's sensor center (figure \ref{fig:angle-scheme}). A detailed description of this process is provided in the companion paper \cite{part1-submittedpaper}. The cameras were shielded with lead blocks to reduce noise from stray radiation. 
 
 {}

\subsection{Dose measurements}
 
 The dosimeter was irradiated with a 6 MV, 600 cGy/min photon beam (Clinac iX, Varian, Palo Alto, USA). The signal-to-noise ratio (SNR) and signal-to-background ratio (SBR) of the detector were studied while varying the dose delivered and the dose rate. Signal-to-noise ratio describes the system's sensitivity and was defined as the ratio of the mean pixel value to its standard deviation for each scintillation spot \cite{lacroix_design_2009}. Signal-to-background was defined as the ratio of the signal to the standard deviation of the background and describes the signal's detectability. 
\begin{equation}
    SNR_{ave} = \frac{\mu_{s}}{\sigma_{s}}, \qquad SNR_{spot}= \sqrt{n}SNR_{ave}, \qquad SBR= \frac{\mu_{spot}}{\sigma_{bg}}
\end{equation}
Different instantaneous dose rates were achieved by varying the distance between the detector and the irradiation source, keeping the delivered monitor units and linac settings constant. 

Each fiber was dose-calibrated by irradiating the phantom with a 6$\times$3 cm$^2$ field size and monitor units (MU) ranging from 3 to 10 MU. The phantom was centered at the isocenter of the linac. Reference dose calculation was performed using a treatment planning system (Raystation; RaySearch laboratories, Stockholm, Sweden). Dose calculations were performed with a 1 mm dose grid. These measurements enabled the light to dose conversion and assessed the linearity of the detector. Then, the developed dosimeter was used to measure the dose distribution and the deformation vector field resulting from a deformation. The dosimeter was imaged and irradiated in both states, i.e. fixed and deformed, with 5$\times$3, 4$\times$3 and 3$\times$3 cm$^2$ field sizes. Dose measurements were validated and compared using an independent scintillation dosimetry system  (Hyperscint; MedScint Inc., Quebec city, Canada).

\begin{figure}
    \centering
    \includegraphics[width = 0.6\textwidth]{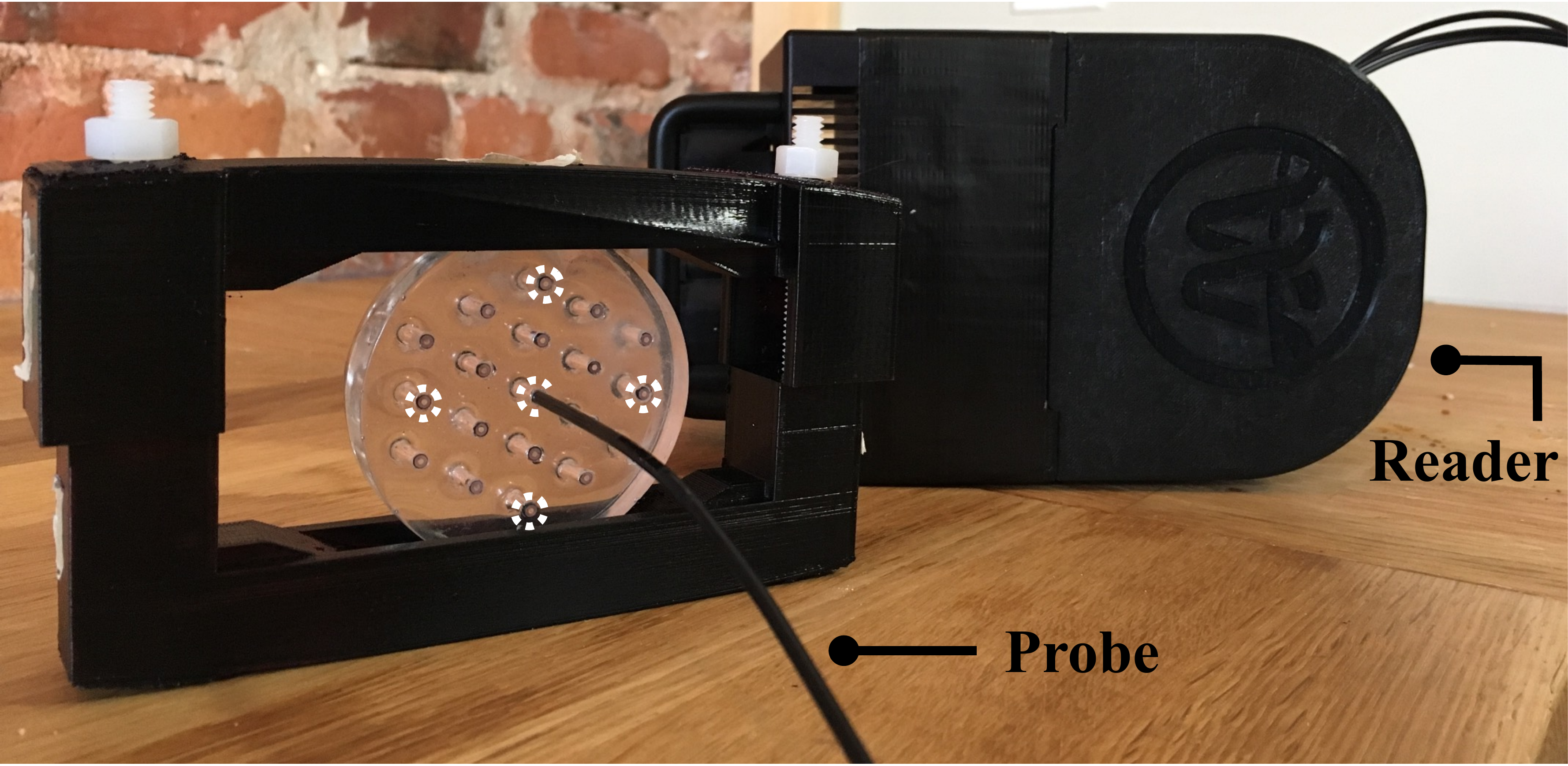}
    \caption{Picture illustrating the Hyperscint customized probe in the deformable dosimeter and its reader. Dose measurements were performed at the location of the five encercled scintillators.}
    \label{fig:hyperscint}
\end{figure}{}

\subsubsection{Independent dose validation}

\begin{figure}
    \centering
    \includegraphics[width = 0.9\textwidth]{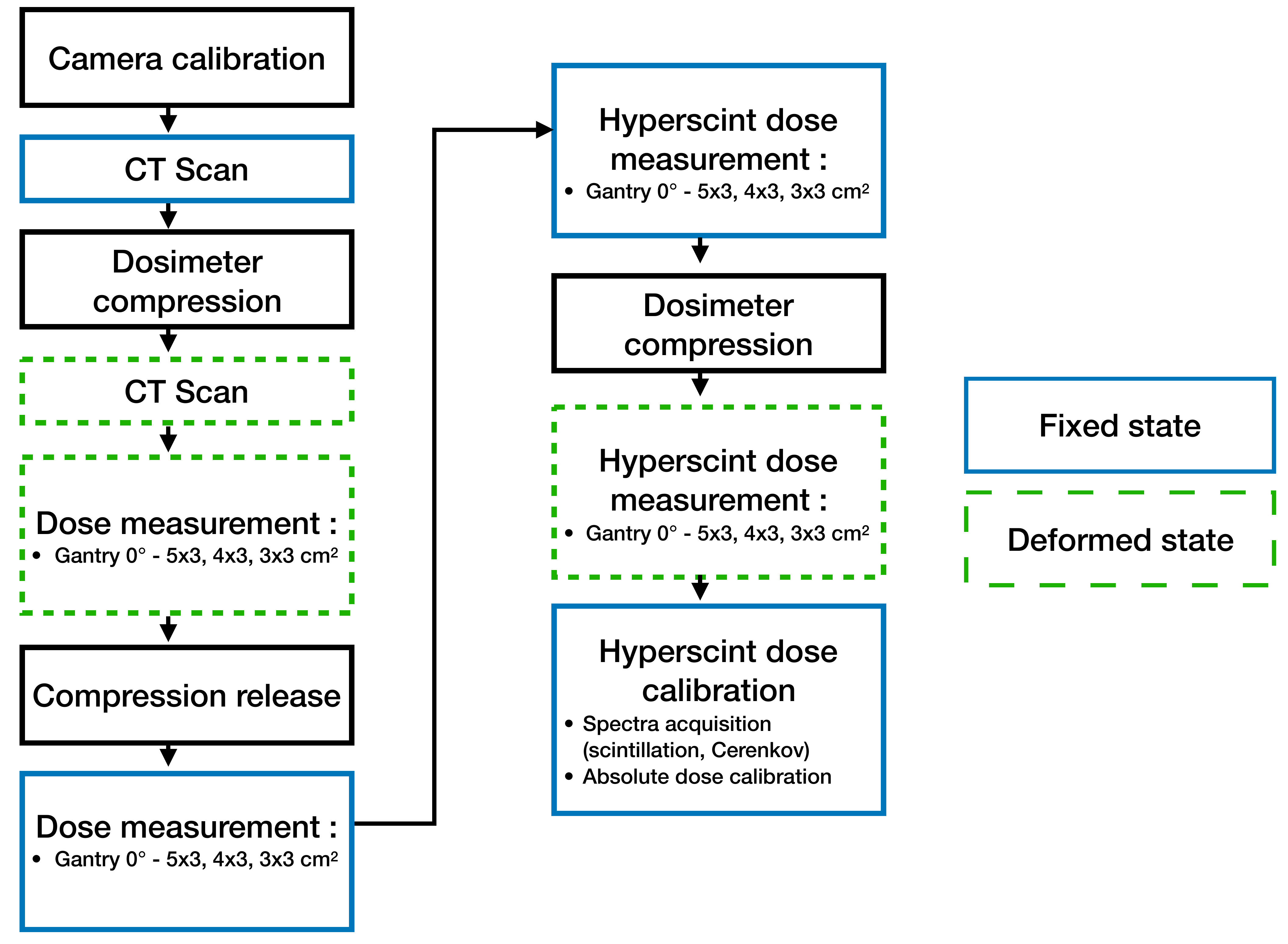}
    \caption{Workflow of the dose measurements and tomographic images acquisition.}
    \label{fig:flowchart}
\end{figure}

Dose measurements previously described were replicated using the Hyperscint scintillation dosimetry research platform. This provided an independent validation of the dose delivered at the location of five chosen scintillators (figure \ref{fig:hyperscint}). A custom manufactured scintillating probe was inserted in the dosimeter at the selected location (replacing the 1.2 cm long scintillator described in section 2.1). The scintillator in the probe has a length and diameter of 1.2 cm and 1 mm respectively, resulting in the same sensitive volume as that of the scintillators used in the deformable dosimeter. The external diameter of the probe matched the internal diameter of the plastic tubing. However in this case, the scintillator was coupled to a 20 m long clear optical fiber to guide the light to a photodetector, thus enabling traditional plastic scintillation dosimetry (PSD) measurements \cite{beaulieu_review_2016}. The system was calibrated at the isocenter of a 10x10 cm$^2$ field, at a depth of 1.5 cm in a solid water phantom (SSD = 98.5 cm). Cerenkov stem signal was corrected using the hyperspectral formalism \cite{archambault_mathematical_2012, therriault-proulx_development_2012}. The scintillation spectrum was measured from a kV irradiation. Cerenkov spectrum was acquired from two MV measurements for which the dose at the scintillator was kept constant: 1) minimal ($C_{min}$), and 2) maximal ($C_{max}$) clear fiber was irradiated in the beam field \cite{guillot_spectral_2011}. The cerenkov spectrum results from the subtraction $C_{max} -C_{min}$. Figure \ref{fig:flowchart} summarizes the workflow of the experimental measurements.

{}

\subsection{Deformation measurements}
Deformation vector fields (DVF) were measured using the dosimeter by tracking the surface centroid of each scintillating fibers, from both sides. Thus, 19 vectors are measured indicating the direction and magnitude of the fiber displacements between the fixed and deformed conditions. Stereo-vision enabled the detection of the 3D position of both fiber ends in the two studied cases (fixed vs deformed). Angulation of the fibers were extracted from the displacement differences measured by the facing stereoscopic pairs. 

The dosimeter was CT-scanned (Siemens Somatom Definition AS Open 64, Siemens Healthcare, Forchheim, Germany), for both conditions. The pitch, current, tube current-time and energy of the scanner were respectively set to 0.35, 60 mA, 1000 mAs and 120 kVp. The CT images were further fed to a DIR algorithm and the computed DVF was extracted. 

The B-Spline algorithm from Plastimatch \cite{sharp_plastimatch_nodate} was used to compute the DVF describing the transformation mapping the fixed dosimeter state to its deformed state. The algorithm's cost function is guided with image similarities using pixel's mean square error (MSE). The regularization term, i.e method to ensure physically realistic deformation scenarios, was set to 0.005. The resulting deformation vector field, obtained optically and from the deformable image registration algorithm, were compared. 

Reproducibility of the deformation and hysteresis of the dosimeter were characterized by tracking the position (fixed and deformed) of the scintillators across 3 deformation repetitions.

\section{Results}
\subsection{Dosimeter calibration and characterization}

\begin{figure}[ht]
    \centering
    \includegraphics[width =0.7\textwidth ]{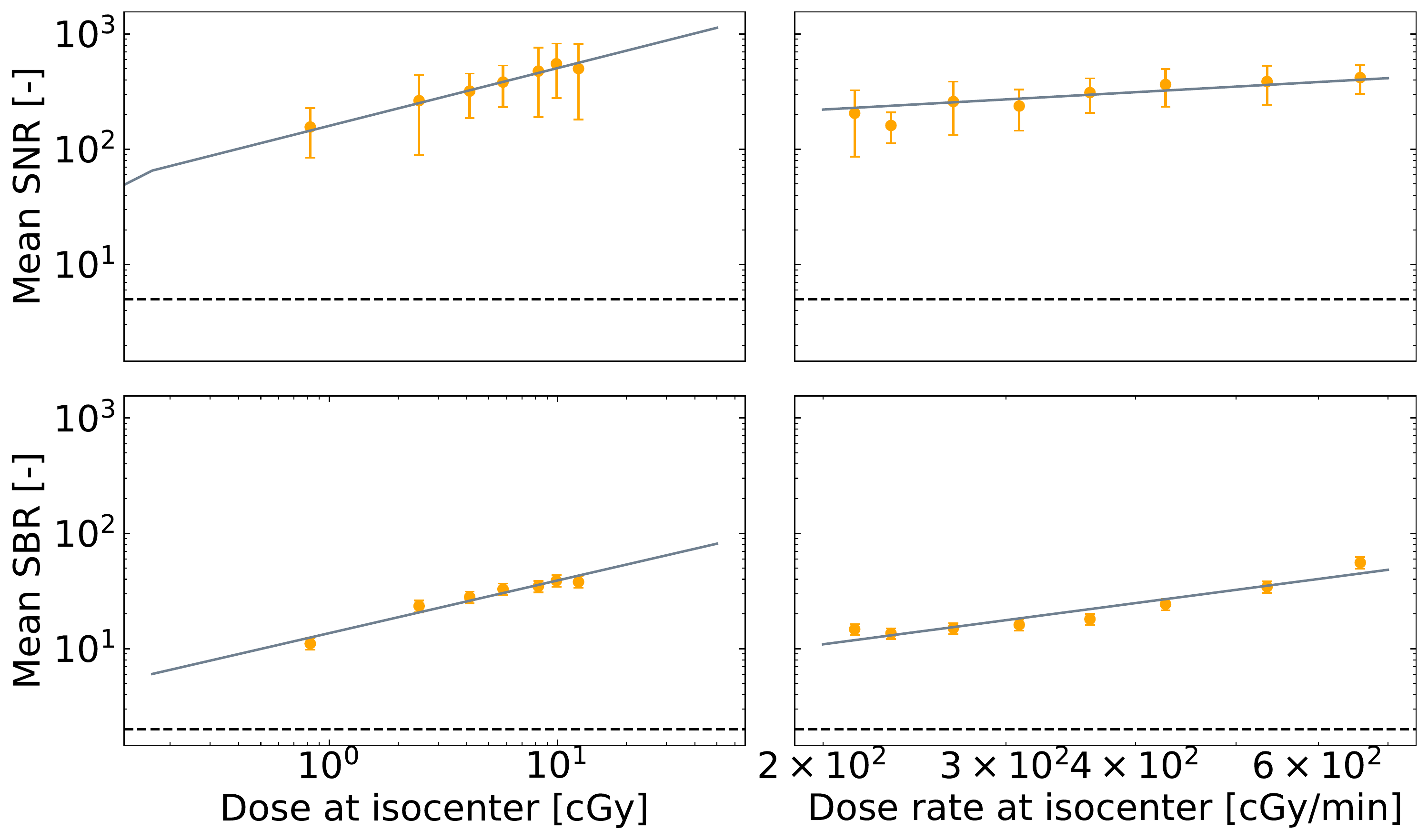}
    \caption{Signal-to-noise ratio (SNR)  and signal-to-background ratio (SBR) as a function of the dose deposited and the dose rate  at the isocenter. Dashed lines represent cut-off values for accurate detectability. Error-bars indicate the range of values obtained for the 19 scintillating fibres rather than the uncertainty on the measure. }
   \label{fig:snr-sbr}
\end{figure}

 Calibration of the detector lead to an expected linear dose-light relationship ($R^2>$ 0.999) for all 19 scintillation fibers. 
For the SNR and SBR analysis, the signal remained over the sensitivity (SNR$>$5) and detectability (SBR$>$2) thresholds for all the explored doses and dose rates (figure \ref{fig:snr-sbr}). Points and error-bars on figure \ref{fig:snr-sbr} represent respectively the mean and standard deviation of the 19 fibers. 
Table \ref{tab: reproducibility} presents the position reproducibility of the 19 scintillators in the fixed and deformed states. Variations in the position of the scintillators (mean $\pm$ standard deviation) are also listed. The higher variations were obtained on the z (depth) axis, but remained under 0.3 mm : the precision of the 3D tracking by the cameras. Hence, the deformation was reproducible and the elastomer did not present hysteresis.

\begin{table}[ht]
\caption{Position variations in the x, y and z axis for the fixed and deformed state over 3 repeated deformations. The table presents mean $\pm$ standard deviations over the 19 points.}
\label{tab: reproducibility}
\centering
\begin{tabular}{l|l|cc}
\br
       &  & Fixed         & Deformed  \\
         \mr
 \multirow{4}{*}{\includegraphics[width = 0.45\textwidth]{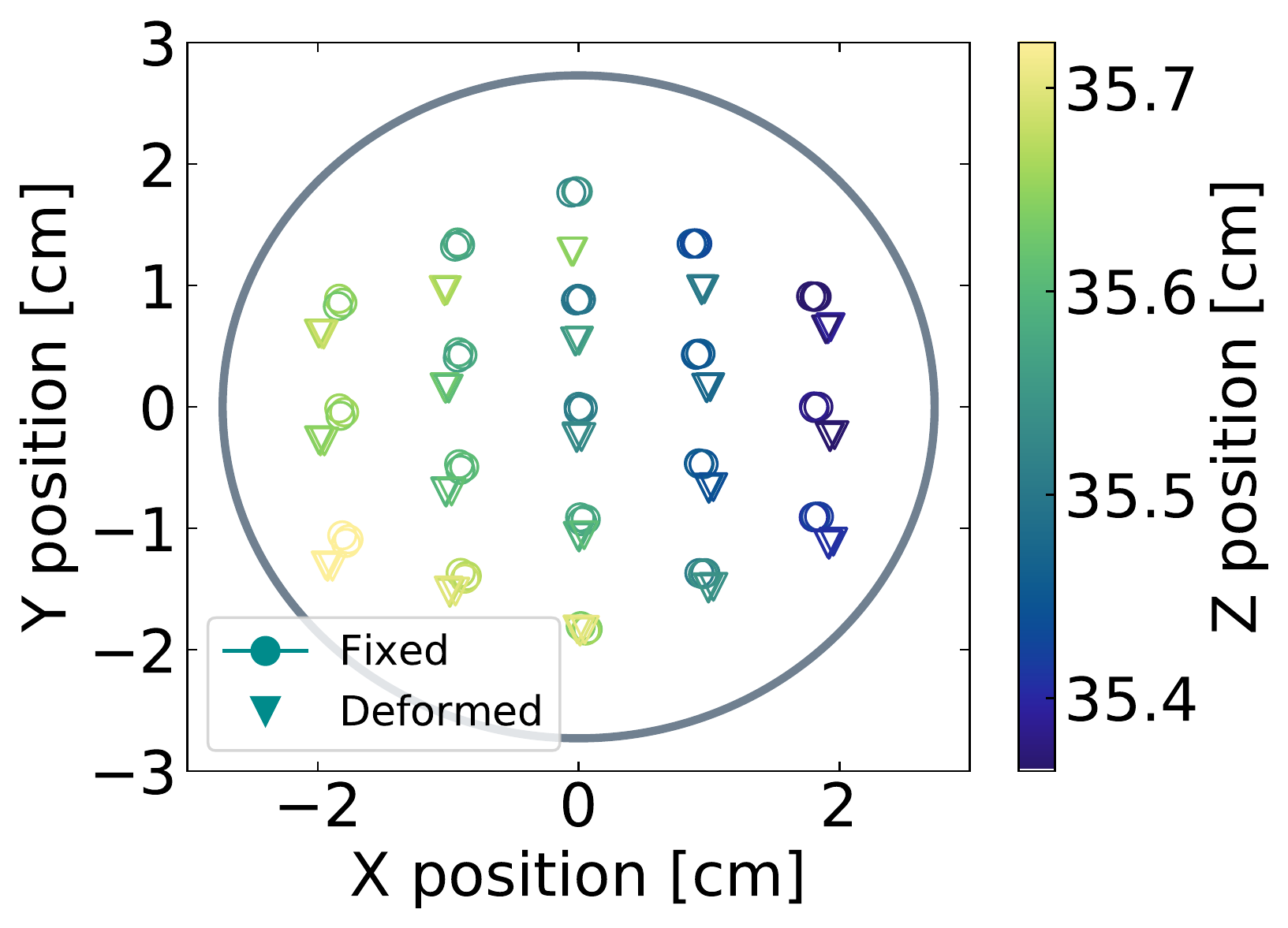} } & x [cm]    & 0.03 $\pm$ 0.01  &-0.009 $\pm$ 0.005 \\
 
 & y [cm] & -0.01 $\pm$ 0.01  & 0.010 $\pm$ 0.006 \\
 & z [cm] &  0.02  $\pm$ 0.01 &0.03 $\pm$ 0.02 \\
 & & & \\
  & & & \\
 & & & \\
& & & \\
 & & & \\
 & & & \\
\br
\end{tabular}
\end{table}

To complete the dosimeter's characterization, a mean density of 1.06$\pm$0.02 g/cm$^2$ was extracted from the CT-scan images, which corroborates its water equivalence.


\subsection{Deformation vector fields}
\begin{figure}[ht]
    \centering
    \includegraphics[width =0.95\textwidth ]{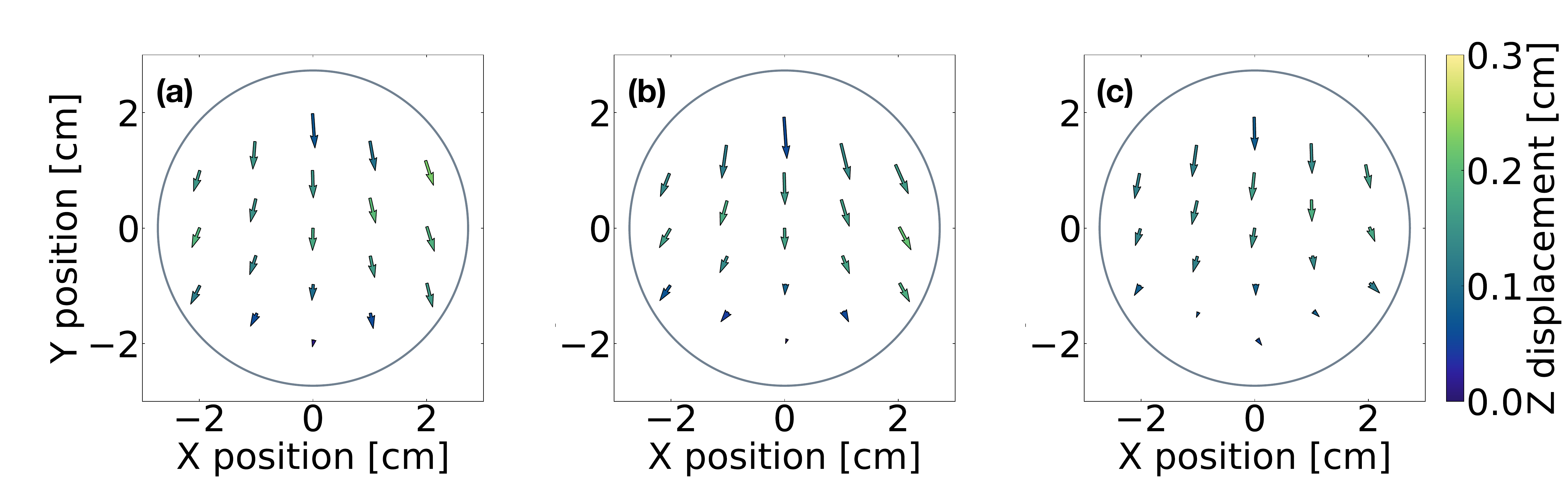}
    \caption{Deformation vector fields measured from scintillation on the first (a) and last (b) dosimeter surfaces, and computed with the Plastimatch DIR algorithm (c). The grey circle represents the dosimeter contour in its fixed state.}
   \label{fig:dvf}
\end{figure}

Figure \ref{fig:dvf} presents the 3D deformation vector fields obtained from the scintillation signal (a, b) and from the Plastimatch deformation algorithm (c). Differences in the DVF measured from both ends (front : figure \ref{fig:dvf}a, back: figure \ref{fig:dvf}b of the elastomer by the facing stereoscopic pairs are attributed to the angulation of the scintillators resulting from the deformation. 

\begin{figure}[ht]
    \centering
    \includegraphics[width =0.8\textwidth ]{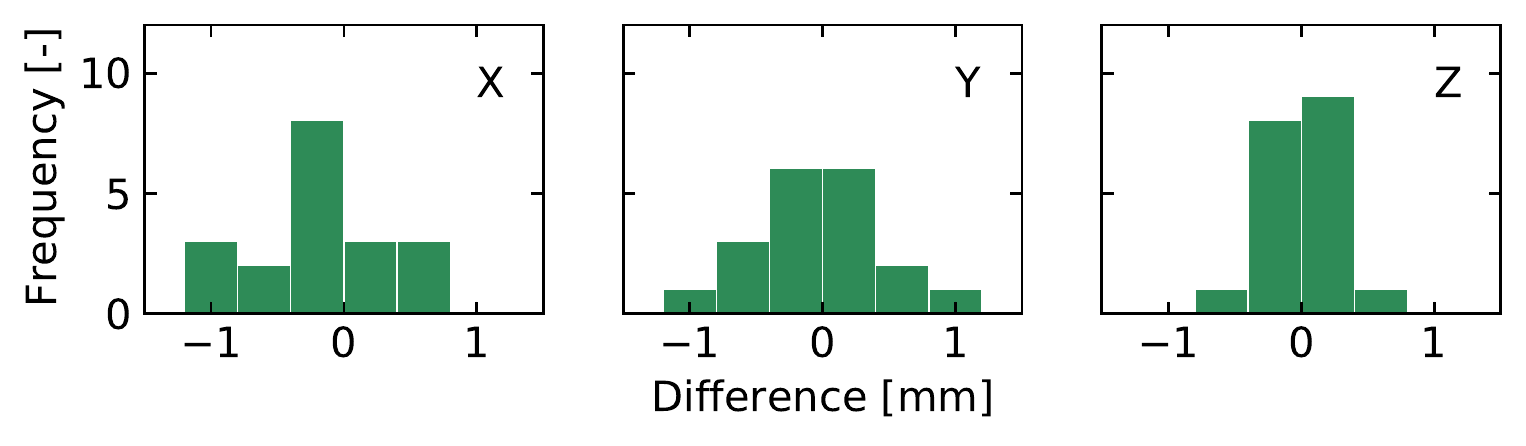}
    \caption{Distribution of differences between the measured and computed DVF in the x, y and z axis, respectively.}
   \label{fig:histo}
\end{figure}

Globally, the DVF computed by the DIR algorithm presents the same shape, and magnitude as the one obtained optically. Overall, the applied compression resulted in a downward shift in the vertical axis and a shift towards the edges in the horizontal axis. Moreover, the compressed dosimeter develop a convex shape towards the cameras (CCD1 and sCMOS1) as a result of the applied deformation. The curve was optically detected by the depth (Z) variation in the 3D tracking. The largest vertical deformation was obtained at the top of the dosimeter with measured and computed displacement of 6.7 $\pm$ 0.6 and 7.1 $\pm$ 0.6 mm.  Figure \ref{fig:histo} presents the distribution of the differences obtained between the measurements from the 3D tracking of the fibers and the ones obtained from the algorithm. Differences remained lower than 1.5 mm and are distributed around 0.

\subsection{Static and deformed dose measurements}

\begin{figure}[ht]
    \centering
    \includegraphics[width =0.7\textwidth ]{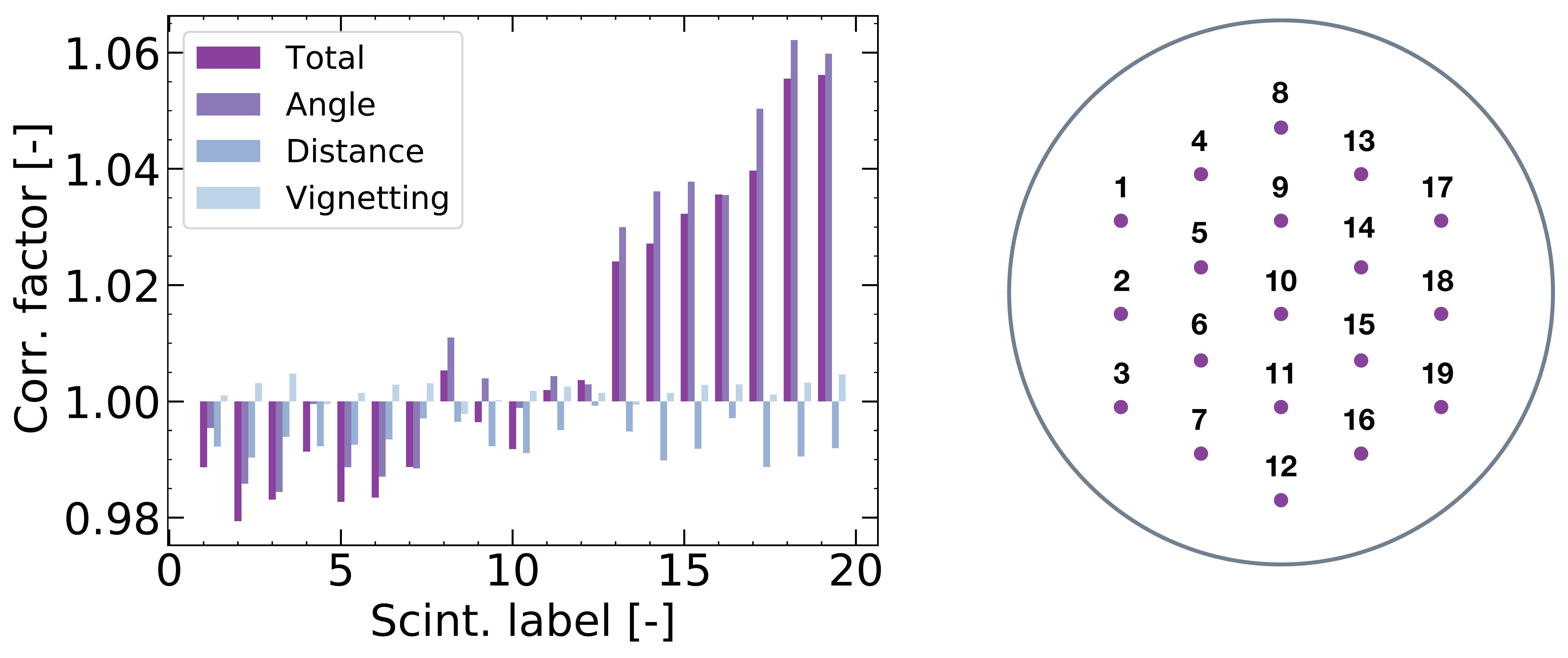}
    \caption{Angular, distal and vignetting correction factors applied to the deformed measurements for each detecting scintillator. The scintillator labels are defined on the right figure.}
   \label{fig:corr_factor}
\end{figure}

\begin{figure}[ht]
    \centering
    \includegraphics[width =0.6\textwidth ]{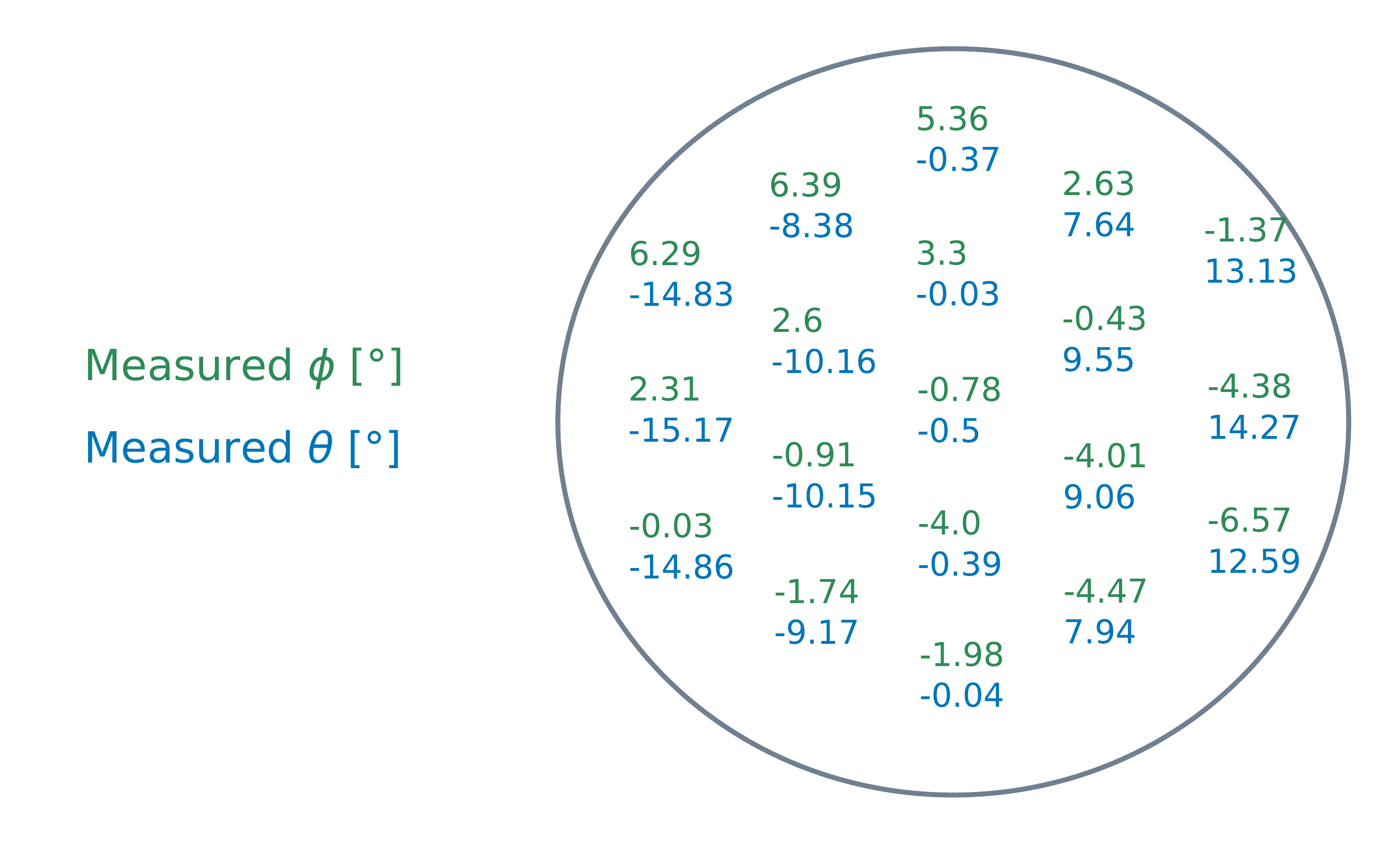}
    \caption{Angulation of the fibers in the $\phi$ and $\theta$ direction.}
   \label{fig:angles}
\end{figure}

Dose distributions were acquired in fixed and deformed conditions. The compression of the deformable dosimeter led to movement, i.e. translation and rotations, of the scintillators. Signal was accordingly corrected to account for variations in the system  collection efficiency. Figure \ref{fig:corr_factor} presents the angular, distal and vignetting corrections that were applied to each scintillator. 
Angulation and distance from CCD1's sensor center were measured. The angulation correction coefficient results from the combined tilt of the scintillators in the elastomer and their position relative to the camera. Deforming the dosimeter led to tilts of the fibers as presented on figure \ref{fig:angles}. Measured $\theta$ presents a symmetry along the x-axis, as expected. 

\begin{figure}[ht]
    \centering
    \includegraphics[width =0.99\textwidth ]{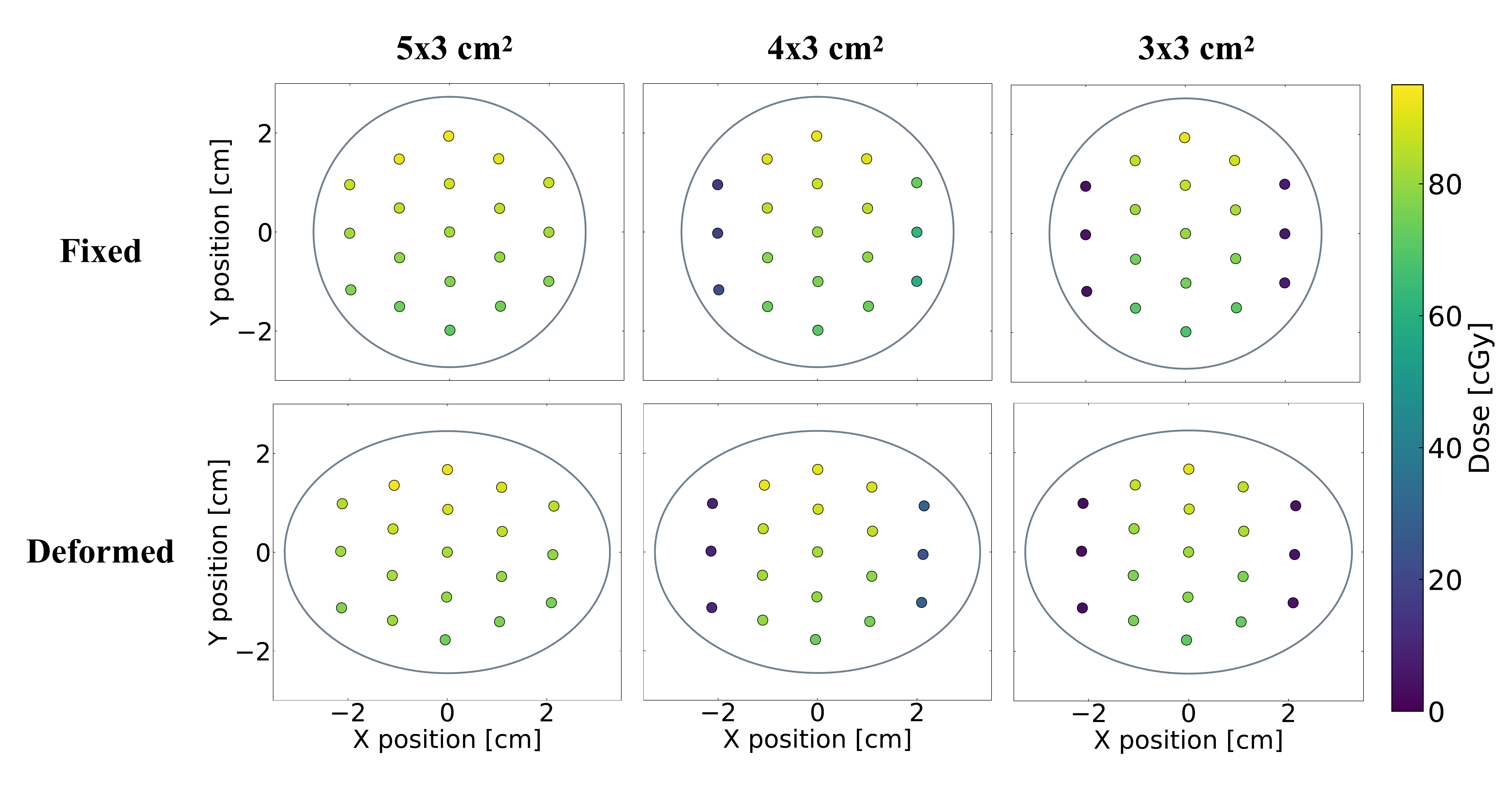}
    \caption{Dose distributions measured from 19 scintillating points for various field sizes in the fixed (top) and deformed (bottom) dosimeter's states. Each colored circle represents the dose measured with a scintillator. Size of the circles is larger than the actual size of scintillators for better visualisation.}
   \label{fig:dose_distribution}
\end{figure}

\begin{figure}[ht]
    \centering
    \includegraphics[width =0.99\textwidth ]{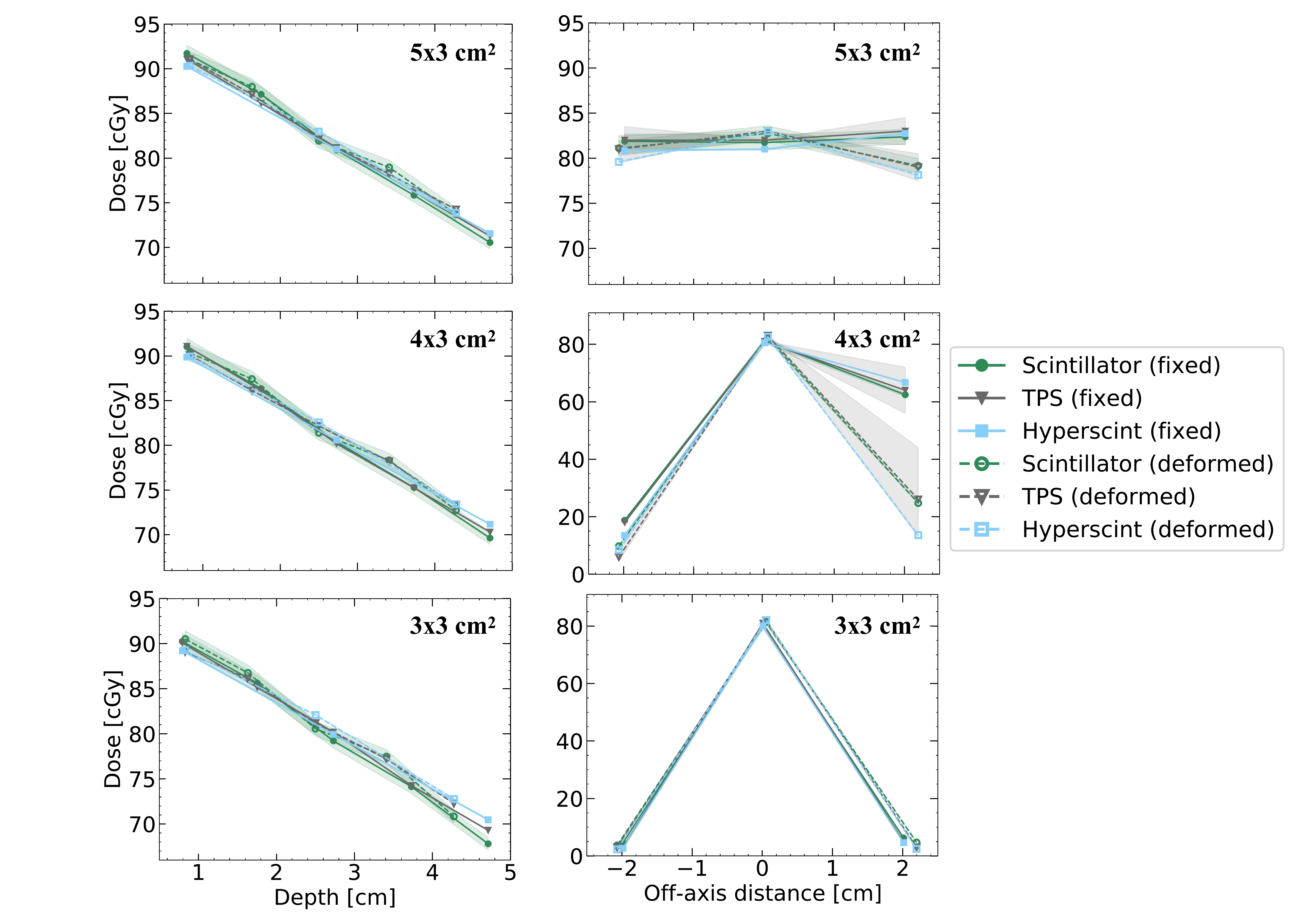}
    \caption{Scintillation dose measurement comparison with calculation from the treatment planning system (TPS) and Hyperscint measurements. Left images present depth dose as right images present dose profiles. }
   \label{fig:pdd_profile}
\end{figure}

Dose distribution from the 19 scintillators are presented on figure \ref{fig:dose_distribution}, for different field sizes. For each field, crossline profiles and depth dose were extracted and compared with the Hyperscint measurements and computation from the treatment planning system (figure \ref{fig:pdd_profile}). An uncertainty of 1\% was estimated on scintillators measurements which mainly takes into account the correction factors uncertainty. Uncertainties on TPS calculations corresponds to dose variations resulting from 1 mm translations to account for setup variations, whereas the uncertainties on the Hyperscint correspond to the standard deviation over 10 consecutive measurements.
For depth dose and profiles, most differences between scintillators measurements and TPS calculations remained within the uncertainty margins of 1\%. In the beam direction, deformation of the dosimeter results in dose shifts along the depth dose line as scintillators were brought closer to the surface.  Scintillators towards the sides of the dosimeter exhibit larger variations between the fixed and deformed conditions. Differences between the fixed and deformed conditions up to 37 cGy (60\%) were obtained, which refers to a scintillator moving through the beam's edge following deformation. It was calculated that differences between the Hyperscint and scintillators measurements, for the 4$\times$3~cm$^2$ profile, were likely caused by a 0.9 mm re-positioning shift, when the probe was inserted in the dosimeter.

\section{Discussion}

We developed a novel real-time deformable dosimeter that can simultaneously measure dose and deformation vector fields with a system of cameras. Using plastic scintillators, we were able to develop a water-equivalent phantom compatible with most imaging modalities. In addition, given the dosimeter's density homogeneity, the scintillators do not act as fiducial markers and allow the evaluation of deformable registration algorithms without influencing their outcomes. However, measuring the light output from displaced scintillators with fixed cameras created new challenges. Hence, it was demonstrated that such system requires precise position and orientation tracking of the scintillators to account for signal variations arising from changes in their optical coupling with the cameras \cite{part1-submittedpaper}. In this case, compressing the dosimeter by 1 cm necessitated correction factors of up to 5.6\%. As displacements of the fibers were lower than 0.71 cm, corrections for angular shifts dominated the total corrections. Altogether, for most of the scintillators, the detecting system measured doses within 0.5 cGy of the TPS calculation and measurements with the Hyperscint system, in both fixed and deformed conditions. The only point presenting a significant discrepancy is the last depth dose point irradiated by the 3$\times$3 cm$^2$ field. The differences between the dose measurements and TPS reached 2.2\%. Other differences between Hyperscint and scintillator measurements remained under the positioning uncertainties. Overall, agreement with the TPS was expected as scintillators were calibrated against calculation from the TPS itself, but with different irradiation conditions than the ones used for analysis. Ideally, the system should be calibrated independently from the TPS. However, dose calibration of the system remains tricky because each detecting scintillator needs to be individually calibrated, to account for variations in the polishing for example, and the phantom doesn't provide sufficient scattering conditions for AAPM TG-51 reference dose calibration \cite{tg51}. As such, using an external dosimetry tool, like a standard PSD dosimeter, to calibrate rather than validate the system would be an interesting avenue.

Deforming the dosimeter with an antero-posterior compression resulted in two main dosimetric effects : 1) along the depth dose, dose to scintillators increased as they were brought closer to the surface and 2) the deformation increased the off-axis distance of scintillators which resulted in dose decrease for scintillators moving from in-field towards the beam penumbra. 4$\times$3 cm$^2$ field profile measurements especially stressed the need for accurate understanding of the deformation as small shift near dose gradients can results in significant dose differences. In that case, 1 mm lateral shifts could result in dose differences up to 40 cGy as the scintillator is close to the beam's edge. The increased complexity of modern radiotherapy techniques, such has IMRT and VMAT type deliveries, further enforces the need for efficient and quantitative dose distribution measurements \cite{low_dosimetry_2019}. Similarly, previous work have demonstrated that a small discrepancy in the computed DVF can significantly impact the warped dose, especially in high gradient regions, highlighting the need for validation \cite{yeo_is_2012}.

AAPM Task Group 132 stated that an ideal DIR validation tools should enable an error detection smaller than the DIR pixel size \cite{brock_use_2017}. In our case, the tomographic images were acquired with an in-plane pixel size of 0.35 mm. The set of stereoscopic pairs of cameras provided an optical measurement of the deformation vector field with a previously demonstrated precision 0.3 mm \cite{part1-submittedpaper}. Hence, the system has the potential to accurately portray deformation vector field for quality assurance applications. The deformation vector field computed with the DIR algorithm presented differences up to 1.5 mm with the one optically measured. AAPM TG-132 stated that an overall registration accuracy within 2 mm is desired for clinical applications \cite{brock_use_2017}. Scan quality, image artifacts and image distortions, amongst others, can affect the resulting quality of a registration. In this work, scan quality was optimized with a tube current-time of 1000 mAs. Contrast was further enhanced by choosing a head scanning protocol. Nevertheless, a computed DVF with Plastimatch present differences from the one predicted by the optical measurements. Those differences are attributed to the known weaknesses of DIR algorithm in homogeneous mediums \cite{yeo_performance_2013}. Hence, DIR algorithms are expected to present lower accuracy in low-contrast regions, such as the dosimeter. 

The prototype developed in this work measured the dose and deformation at 19 points. However, it is to be stated that the scintillators number and density could easily be increased for the need of an aimed application. Moreover, the shape and size of the elastomer is solely limited by the mold it is cast in. Hence, the dosimeter's design is customizable. Moreover, due to its water equivalence, the phantom is compatible with most imaging modalities namely CT, CBCT and MRI. Given the demonstrated advantages of scintillators, a deformable scintillator-based dosimeter would be well-suited to the development of anthropomorphous phantoms to further investigate DIR validation. Thus, future work will look at the development of a dosimeter comprising different density regions to better mimic tissues and tumor. 

The developed dosimeter relies on image acquisitions from four perspectives. Using a set of four cameras, accurate correction of translations and rotations of the scintillators following a deformation is achievable. Yet, such a system comes with an increase complexity regarding the calibration and acquisition of the image sets. 2D and 3D scintillation dosimetry have previously been shown feasible using a single camera \cite{goulet_novel_2014, alexander_1_nodate}. However, the proposed stereoscopic system enables robust measurements that allows deforming and moving the dosimeter to mimic anatomical variations. Moreover, given the recent emergence of low-cost CMOS and new generation of CCDs, the increased number of photodetectors should not limit the clinical implementation of such a system.

\section{Conclusion}

 Anatomical motion and deformation challenge the calculation of the dose delivered, raising the need for adapted quality assurance tools. We developed a dosimeter that enables measurements in fixed and deformable conditions, while tracking the deformation itself. The water-equivalent composition of the dosimeter further endows it with the quality to act both as a phantom and detector.  Moreover, the detector allows a wide variety of 2D and 3D geometric or anthropomorphous designs since its shape and size is solely determined by the mold used to cast the elastomer. Such a detector could be used for the quality assurance of DIR algorithms and to explore the dosimetric impact of organ deformations.

\section{Acknowledgement} 
We thank Medscint, especially Benjamin Côté and Simon Lambert-Girard, for their support and for kindly providing a customized probe as well as the Hyperscint research platform for the measurements. We also thank Jonathan Boivin and Ève Chamberland for their assistance in CT image acquisition and dose calculations, respectively. This work was financed by the Natural Sciences and Engineering Research Council of Canada (NSERC) Discovery grants \#2019-05038 and \#2018-04055. Emily Cloutier acknowledges support by the Fonds de Recherche du Quebec – Nature et Technologies (FRQNT). The authors thank Ghyslain Leclerc for the English revision of the paper.


\begin{small}
\printbibliography[heading=bibintoc]
\end{small}

\end{document}